\documentclass[onecolumn,showpacs,preprintnumbers]{revtex4}
\usepackage{amsthm, amscd, amsfonts, amssymb, graphicx}
\usepackage{dcolumn}
\usepackage{bm}
\usepackage[english]{babel}

\begin{document}

\title{Toy model of superconductivity.}
\author{Konstantin V. Grigorishin}
\email{gkonst@ukr.net}
\author{Bohdan I. Lev}
\email{bohdan.lev@gmail.com} \affiliation{Boholyubov Institute for
Theoretical Physics of the National Academy of Sciences of
Ukraine, 14-b Metrolohichna str. Kiev-03680, Ukraine.}
\date{\today}

\begin{abstract}
The model of hypothetical superconductivity, where the energy gap
asymptotically approaches zero as temperature or magnetic field
increases, has been proposed. Formally the critical temperature
and the second critical field for such a superconductor is equal
to infinity. Thus the material is in superconducting state always.
\end{abstract}


 \maketitle


Critical temperature $T_{\texttt{C}}$ and critical magnetic fields
$H_{\texttt{c}}$, $H_{\texttt{c2}}$ are most important
characteristics of a superconductor. The critical parameters
depends on an effective coupling constant with some collective
excitations $g=\nu_{F}\lambda\lesssim 1$ (here $\nu_{F}$ is a
density of states at Fermi level, $\lambda$ is an interaction
constant), on frequency of the collective excitations $\omega$ and
on correlation length $\xi_{0}$. The larger coupling constant, the
larger these critical parameters. For example, for large values of
$g$ we have $T_{\texttt{C}} \propto \omega\sqrt{g}$
\cite{mahan,ginz} (or $T_{\texttt{C}} \propto \omega g$ in BCS
theory). Formally the critical temperature can be made arbitrarily
large by increasing the electron-phonon coupling constant. However
in order to reach room temperature such values of the coupling
constant are necessary which are not possible in real materials.
Moreover we can increase the frequency $\omega$ due nonphonon
pairing mechanisms as proposed in \cite{ginz}. However with
increasing of the frequency the coupling constant decreases as
$g\propto 1/\omega$, therefore
$T_{\texttt{C}}(\omega\rightarrow\infty)=1.14\omega\exp\left(-1/g\right)\rightarrow
0$. The second critical magnetic field can be enlarge due to the
decrease of the correlation length in "dirty limit"
$\xi=\sqrt{\xi_{0}l}$ \cite{sad}, where $l$ is a free length.
However the critical field is low near the critical temperature:
$H_{\texttt{c2}}(T\rightarrow T_{\texttt{C}})\rightarrow 0$. In a
present work we generalize BCS model so that the problem of the
critical parameters is removed due to the fact that a ratio
between the gap and the critical temperature
($2\Delta/T_{\texttt{C}}=3\div 7$ for presently known materials)
is changed to $2\Delta/T_{\texttt{C}}\rightarrow 0$. We consider a
system of fermions with Hamiltonian:
\begin{eqnarray}\label{2.1}
    \widehat{H}=\sum_{\textbf{k},\sigma}\xi(k)a_{\textbf{k},\sigma}^{+}a_{\textbf{k},\sigma}
    -\frac{\lambda}{V}\sum_{\textbf{k},\textbf{p}}a_{\textbf{p}\uparrow}^{+}a_{-\textbf{p}\downarrow}^{+}a_{-\textbf{k}\downarrow}a_{\textbf{k}\uparrow}
    +\upsilon\sum_{\textbf{k}}\left[\frac{\Delta}{|\Delta|}a_{\textbf{k}\uparrow}^{+}a_{-\textbf{k}\downarrow}^{+}
    +\frac{\Delta^{+}}{|\Delta|}a_{-\textbf{k}\downarrow}a_{\textbf{k}\uparrow}\right]
    \equiv \widehat{H}_{\texttt{BCS}}+\widehat{H}_{\texttt{\texttt{ext}}},
\end{eqnarray}
where $\widehat{H}_{\texttt{BCS}}$ is BCS Hamiltonian - kinetic
energy + pairing interaction ($\lambda>0$), energy $\xi(k)\approx
v_{F}(|\textbf{k}|-k_{F})$ is counted from Fermy surface. The term
$\widehat{H}_{\texttt{ext}}$ is the external pair potential or
"source term" \cite{matt1}. Operators
$a_{\textbf{k}\uparrow}^{+}a_{-\textbf{k}\downarrow}^{+}$ and
$a_{-\textbf{k}\downarrow}a_{\textbf{k}\uparrow}$ are creation and
annihilation of Cooper pair operators \cite{schr}, $\Delta$ and
$\Delta^{+}$ are anomalous averages:
\begin{eqnarray}\label{2.2}
    \Delta^{+}=\frac{\lambda}{V}\sum_{\textbf{p}}\left\langle
    a_{\textbf{p}\uparrow}^{+}a_{-\textbf{p}\downarrow}^{+}\right\rangle,
    \quad
    \Delta=\frac{\lambda}{V}\sum_{\textbf{p}}\left\langle
    a_{-\textbf{p}\downarrow}a_{\textbf{p}\uparrow}\right\rangle,
\end{eqnarray}
which are the complex order parameter
$\Delta=|\Delta|e^{i\theta}$. The multipliers
$\frac{\Delta}{|\Delta|}$ and $\frac{\Delta^{+}}{|\Delta|}$ are
introduced into $\widehat{H}_{\texttt{ext}}$ in order that the
energy does not depend on the phase $\theta$ ($a\rightarrow
ae^{i\theta/2},a^{+}\rightarrow a^{+}e^{-i\theta/2}\Longrightarrow
\Delta\rightarrow\Delta e^{i\theta},\Delta^{+}\rightarrow
\Delta^{+}e^{-i\theta}$). Thus both $\widehat{H}_{\texttt{BCS}}$
and $\widehat{H}_{\texttt{ext}}$ is invariant under the $U(1)$
transformation unlike the source term in \cite{matt1} where it has
a noninvariant form
$\upsilon\sum\left[a_{\textbf{k}\uparrow}^{+}a_{-\textbf{k}\downarrow}^{+}
+a_{-\textbf{k}\downarrow}a_{\textbf{k}\uparrow}\right]$. Hence
$\upsilon$ is energy of a Cooper pair relative to uncoupled state
of the electrons in the external pair potential
$H_{\texttt{ext}}$. It should be noted that the energy gap
$|\Delta|$ is energy of a Cooper pair relative to uncoupled state
of the electrons too. However the field $\Delta$ is a
self-consistent field as a consequence of attraction between
electrons. The field $\upsilon$ is the applied field to the system
from the outside.

\begin{figure}[ht]
\includegraphics[width=8.5cm]{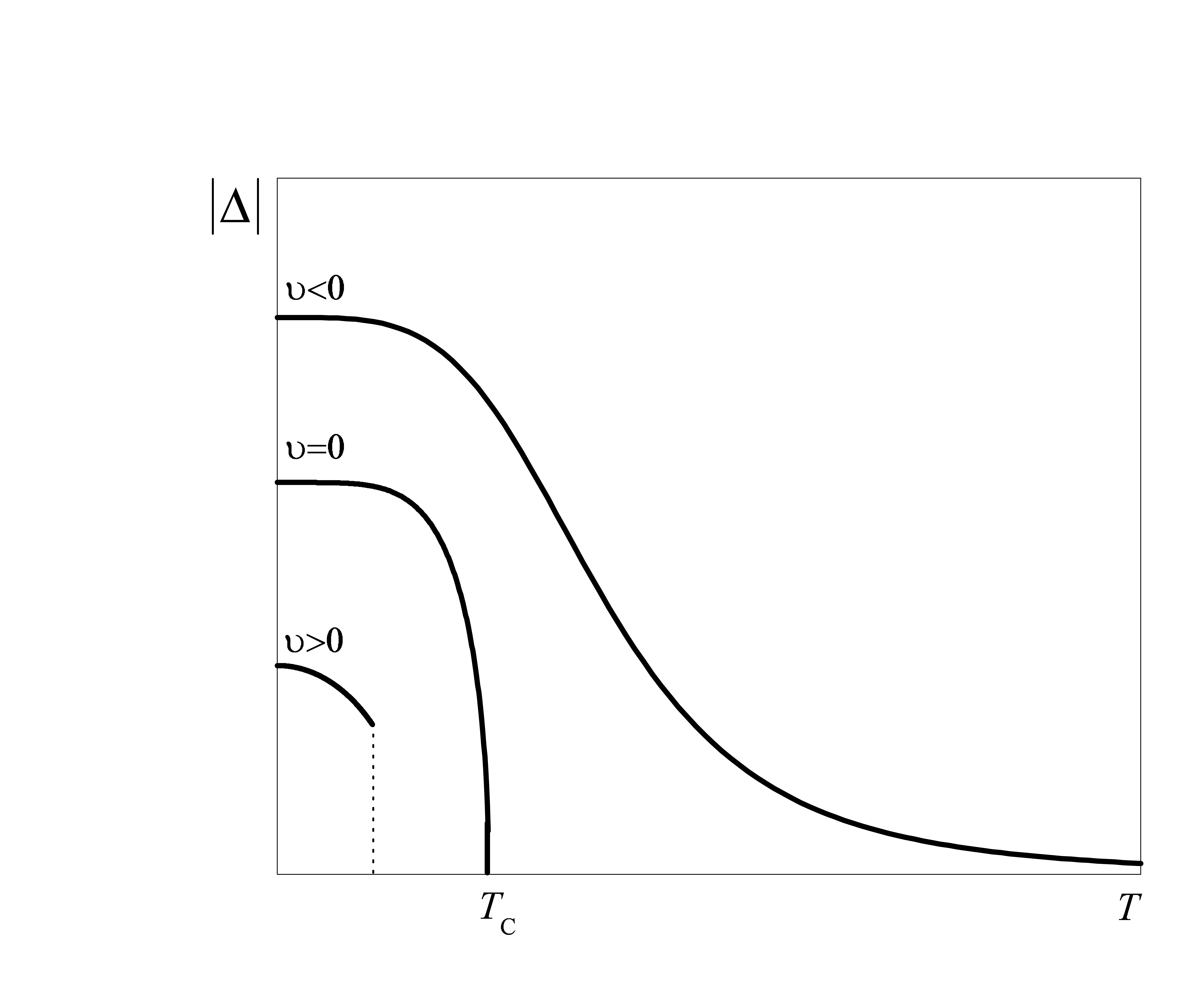}
\caption{Energy gaps $\Delta(T)$ as solution of Eq.(\ref{2.6}) for
three values of the external pair potential $\upsilon$.}
\label{Fig1}
\end{figure}

Using the Fermi commutation relations and the anomalous averages
(\ref{2.2}), Hamiltonian (\ref{2.1}) can be rewritten in a form
\begin{eqnarray}\label{2.3}
    \widehat{H}=\sum_{\textbf{k},\sigma}\xi(k)a_{\textbf{k},\sigma}^{+}a_{\textbf{k},\sigma}
    +\left(1-\frac{\upsilon}{|\Delta|}\right)\sum_{\textbf{k}}\left[\Delta^{+}a_{\textbf{k}\uparrow}a_{-\textbf{k}\downarrow}
    +\Delta a_{-\textbf{k}\downarrow}^{+}a_{\textbf{k}\uparrow}^{+}\right]+\frac{1}{\lambda}V|\Delta|^{2}.
\end{eqnarray}
Then normal $G$ and anomalous $F$ propagators have forms:
\begin{eqnarray}
    G=i\frac{i\varepsilon_{n}+\xi}
    {(i\varepsilon_{n})^{2}-\xi^{2}-|\Delta|^{2}(1-\upsilon/|\Delta|)^{2}}\label{2.5a}\\
    F=i\frac{\Delta(1-\upsilon/|\Delta|)}
    {(i\varepsilon_{n})^{2}-\xi^{2}-|\Delta|^{2}(1-\upsilon/|\Delta|)^{2}},\label{2.5b}
\end{eqnarray}
where $\varepsilon_{n}=\pi T(2n+1)$ \cite{matt2}. Then from
Eq.(\ref{2.2}) we have self-consistency condition for the order
parameter
\begin{equation}\label{2.6}
    \Delta=\lambda \nu_{F}
    T\sum_{n=-\infty}^{\infty}\int_{-\omega}^{\omega}d\xi iF(\varepsilon_{n},\xi)
    \Longrightarrow 1=g\int_{-\omega}^{\omega}d\xi
    \frac{1-\upsilon/|\Delta|}{2\sqrt{\xi^{2}+|\Delta|^{2}(1-\upsilon/|\Delta|)^{2}}}
    \tanh\frac{\sqrt{\xi^{2}+|\Delta|^{2}(1-\upsilon/|\Delta|)^{2}}}{2T}.
\end{equation}
Solutions of Eq.(\ref{2.6}) are shown in Fig.{\ref{Fig1}}. If the
external pair potential is absent $\upsilon=0$ we have usual
self-consistency equation for the gap $\Delta$: the gap is a
function of temperature such that $\Delta(T\geq
T_{\texttt{C}})=0$. The larger the coupling constant
$g=\lambda\nu_{F}$, the larger $T_{\texttt{C}}$. If $\upsilon>0$
then the pairing of quasiparticles results in increase of the
system's energy that suppresses superconductivity and first order
phase transition takes place. If $\upsilon<0$ then the pairing
results in decrease of the system's energy. In this case a
solution of Eq.(\ref{2.6}) is such that the gap $\Delta$ does not
vanish at any temperature. At large temperature $T\gg
T_{\texttt{C}}$ the gap is
\begin{equation}\label{2.7}
    |\Delta(T\rightarrow\infty)|=\frac{g\omega|\upsilon|}{2T}.
\end{equation}
Then the critical temperature is $T_{\texttt{C}}=\infty$ (in
reality it limited by the melting of the substance). It should be
noted that if $\lambda=0$ then for any $\upsilon$ a
superconducting state does not exist ($\Delta=0$ always). This
means electron-electron coupling is the cause of the transition to
superconducting state only but not the external pair potential
$\upsilon$.

\begin{figure}[ht]
\includegraphics[width=8.5cm]{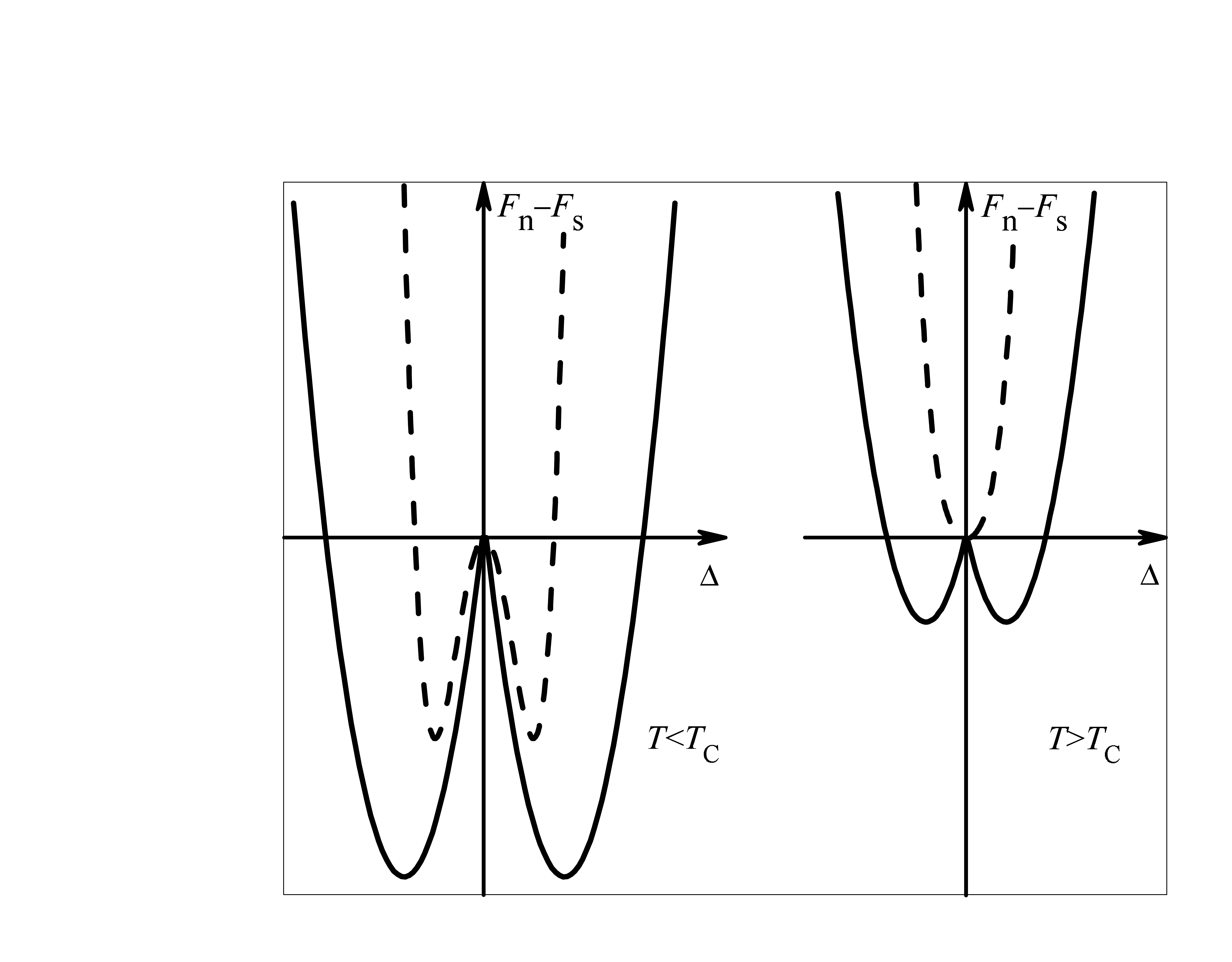}
\caption{Free energies Eq.(\ref{3.1}) (dash line) and
Eq.(\ref{3.3}) (solid line) at $T<T_{\texttt{C}}$ and
$T>T_{\texttt{C}}$ ($q=0$) }
 \label{Fig2}
\end{figure}

For investigation of thermodynamic and electrodynamic properties
of the system we should to find a free energy. Greatest interest
is the case $\upsilon<0$ in a limit $T\rightarrow\infty$. We can
see in Fig.(\ref{Fig1}) that $\Delta$ ia small in this region.
This means that as a starting point we can take the Landau
expansion (in a momentum space) \cite{sad}:
\begin{equation}\label{3.1}
    F_{\texttt{s}}=F_{\texttt{n}}+A|\Delta|^{2}+\frac{B}{2}|\Delta|^{4}+q^{2}C|\Delta|^{2},
\end{equation}
where
\begin{equation}\label{3.2}
    A=\nu_{F}\frac{T-T_{\texttt{C}}}{T_{\texttt{C}}},\quad
    B=\nu_{F}\frac{7\zeta(3)}{8\pi^{2}T_{\texttt{C}}^{2}},\quad
    C=\nu_{F}\xi_{0}^{2},
\end{equation}
$\textbf{q}$ is momentum of a Cooper pair, $\xi_{0}$ is a
coherence length at $T=0$, $F_{\texttt{n}}$ is a free energy of a
normal state. In a limit $T\gg T_{\texttt{C}}$ we can write a
coefficient $A$ as $A=\nu_{F}T/\widetilde{T}_{\texttt{C}}>0$,
where $\widetilde{T}_{\texttt{C}}$ is an adjustable parameter now,
and we should to add to the free energy a term
$\langle\widehat{H}_{\texttt{\texttt{ext}}}\rangle$. Using the the
anomalous averages (\ref{2.2}) we can obtain
$\langle\widehat{H}_{\texttt{\texttt{ext}}}\rangle=\frac{2\upsilon}{\lambda}|\Delta|<0$.
Then we the free energy has a form
\begin{equation}\label{3.3}
    F_{\texttt{s}}=F_{\texttt{n}}+A|\Delta|^{2}+q^{2}C|\Delta|^{2}+\frac{2\upsilon}{\lambda}|\Delta|.
\end{equation}
A term $\frac{B}{2}|\Delta|^{4}$ can be omitted due to the
smallness of the gap. Minimization of the free energy with respect
to $|\Delta|$ (if $q=0$) gives:
\begin{equation}\label{3.4}
    |\Delta|=\frac{|\upsilon|}{A\lambda}=\frac{g\omega|\upsilon|}{2T}\quad\Rightarrow\quad
    A=\frac{2T}{\nu_{F}\lambda^{2}\omega},
\end{equation}
where we must assume
$\widetilde{T}_{\texttt{C}}=\frac{g^{2}\omega}{2}$ in order to get
Eq.(\ref{2.7}). Difference of the free energy (\ref{3.1}) from the
free energy (\ref{3.4}) is shown in Fig.\ref{Fig2}. We can see
that at $\upsilon<0$ a superconducting phase exists at any
temperature.
Then in coordinate space we can write Gibbs free energy as
\begin{equation}\label{3.5}
    G_{\texttt{s}}=G_{\texttt{n}}+a|\Psi|^{2}+2u|\Psi|+\frac{1}{4m}\left|\left(-i\hbar\nabla-\frac{2e}{c}\textbf{A}\right)\Psi\right|^{2}
    +\frac{H^{2}}{8\pi}-\frac{\textbf{HH}_{0}}{4\pi},
\end{equation}
where $\textbf{H}$ is a microscopic magnetic field in each point
of a superconductor, $\textbf{H}_{0}$ is strength of an external
homogeneous magnetic field, $\textbf{A}=\texttt{rot}\textbf{H}$ is
a vector-potential. Coefficient $a$ is proportional to temperature
$a=\alpha T>0$, coefficient $u$ is proportional to the external
pair potential $u=\eta\upsilon<0$. The Eq.(\ref{3.5}) is valid at
$T\gg T_{\texttt{C}}$ only. For regions $T\sim T_{\texttt{C}}$ and
$T\rightarrow 0$ we must replace $\alpha T\rightarrow\alpha
(T-T_{\texttt{C}})$ and take into account a term $b/2|\Psi|^{4}$.

The free energy (\ref{3.5}) can be made dimensionless:
\begin{equation}\label{3.6}
    G_{\texttt{s}}=G_{\texttt{n}}+\frac{H^{2}_{\texttt{c}}}{8\pi}\left[|\varphi|^{2}-2|\varphi|+
    \xi^{2}\left|\left(i\nabla+\frac{2\pi}{\Phi_{0}}\textbf{A}\right)\varphi\right|^{2}\right]
    +\frac{H^{2}}{8\pi}-\frac{\textbf{HH}_{0}}{4\pi}.
\end{equation}
Then Lagrange equations are
\begin{eqnarray}
      &&\xi^{2}|\varphi|\left(i\nabla+\frac{2\pi}{\Phi_{0}}\textbf{A}\right)^{2}\varphi+\varphi|\varphi|-\varphi=0 \label{3.7a}\\
      \nonumber\\
      &&\texttt{rot}\texttt{rot}\textbf{A}
      =-i\frac{\Phi_{0}}{4\pi\lambda^{2}}\left(\varphi^{+}\nabla\varphi-\varphi\nabla\varphi^{+}\right)
      -\frac{|\varphi|^{2}}{\lambda^{2}}\textbf{A},\label{3.7b}
\end{eqnarray}
and a boundary condition are
\begin{equation}\label{3.8}
\left(i\hbar\nabla+\frac{2\pi}{\Phi_{0}}\textbf{A}\right)\textbf{n}\varphi=0.
\end{equation}
Here $\varphi$ is a dimensionless order parameter and
$H_{\texttt{c}}$ is a critical magnetic field:
\begin{eqnarray}
    &&|\Psi\left(\textbf{A}=0,\nabla\Psi=0\right)|\equiv\Psi_{0}=\frac{|u|}{a}\sim\frac{\upsilon}{T}\Longrightarrow
    \varphi=\frac{\Psi}{\Psi_{0}}\label{3.8}\\
    &&\frac{H_{\texttt{c}}^{2}}{8\pi}=\frac{u^{2}}{a}\Longrightarrow H_{\texttt{c}}\sim\frac{|\upsilon|}{\sqrt{T}},\label{3.9}
\end{eqnarray}
$\textbf{n}$ is a normal to superconductor's surface,
$\Phi_{0}=\pi\hbar c/e$ is a the magnetic flux quantum.
Correlation length $\xi$, magnetic field penetration depth
$\lambda$ and Ginzburg-Landau parameter $\chi$ are
\begin{eqnarray}
  &&\xi^{2} = \frac{\hbar^{2}}{4ma}\Longrightarrow\xi\sim\frac{1}{\sqrt{T}} \label{3.10}\\
  &&\frac{1}{\lambda^{2}} = \frac{8\pi e^{2}}{mc^{2}}|\Psi_{0}|^{2}\Longrightarrow\lambda\sim\frac{T}{|\upsilon|} \label{3.11}\\
  &&\chi = \frac{\lambda}{\xi}\sim\frac{T^{3/2}}{|\upsilon|} \label{3.12}
\end{eqnarray}
The proportionality of the penetration depth to the temperature
and the inverse proportionality to the external pair potential -
Eq.(\ref{3.11}) is the expected result. Greater attention should
be given to a reduction of the correlation length with temperature
- Eq.(\ref{3.10}). We can see $\xi$ is determined by properties of
a superconductor only (at $T\gg T_{\texttt{C}}$). We know that the
correlation length depends on temperature as
$\xi=\xi_{0}/\sqrt{|1-T/T_{\texttt{C}}|}$. That is at
$T<T_{\texttt{C}}$ it increases with increasing temperature, at
$T=T_{\texttt{C}}$ it diverges, at $T>T_{\texttt{C}}$ it decreases
with increasing temperature as $1/T$ (at large $T$). However above
the critical temperature the correlation length has physical sense
of the size of a superconducting phase nucleus in a normal
conductor. Superconducting phase at $T>T_{\texttt{C}}$ is
energetically unfavorable and because it arises fluctuationally by
bubble size $\xi$. Switching of the field $\upsilon$ changes the
situation. The field holds fluctuationally arisen superconducting
phase nucleuses. This continues until the superconducting phase
does not fill the entire volume of the metal. From Eq.(\ref{3.12})
we can see the Ginzburg-Landau parameter increases with
temperature as $T^{3/2}$ unlike usual superconductors where the
parameter is constant. This means that at large temperature all
superconductors in the external pair field become type II
superconductors.

Besides the critical temperature important characteristics of a
superconductor are the first $H_{\texttt{c1}}$ and the second
$H_{\texttt{c2}}$ critical fields. The first critical field is
half as much than a field of single vortex which can be determined
from Eq.(\ref{3.7b}). Thus we have
\begin{equation}\label{3.13}
    H_{\texttt{c1}}=\frac{\Phi_{0}}{4\pi\lambda^{2}}\ln\chi\sim\frac{\upsilon^{2}}{T^{2}}
\end{equation}
Hence critical current of emergence of resistance is
\begin{equation}\label{3.13a}
    I_{\texttt{c1}}=\frac{1}{2}H_{\texttt{c1}}cR\sim\frac{\upsilon^{2}}{T^{2}},
\end{equation}
where $R$ is radius of a wire \cite{tinh}. For calculation of
$H_{\texttt{c2}}$ we can use the method presented in Appendix
\ref{field}. Then Eq.(\ref{3.7b}) has a form
\begin{equation}\label{3.14}
    \xi^{2}|\varphi|\left[-\frac{\texttt{d}^{2}}{\texttt{d}x^{2}}+\frac{2\pi i}{\Phi_{0}}Hx\frac{\texttt{d}}{\texttt{d}y}
    +\left(\frac{2\pi
    H}{\Phi_{0}}\right)^{2}x^{2}\right]\varphi+\varphi|\varphi|-\varphi=0
\end{equation}
We can consider the order parameter is real $\varphi=\varphi^{+}$
and average it over the system so that
$\langle\varphi(x,y)\rangle=\texttt{const}=\varphi>0$. Then we
have
\begin{equation}\label{3.15}
    \xi^{4}\left(\frac{2\pi
    H}{\Phi_{0}}\right)^{2}\varphi+\varphi-1=0,
\end{equation}
and the order parameter is
\begin{equation}\label{3.16}
    \varphi=\frac{1}{1+\xi^{4}\left(\frac{2\pi H}{\Phi_{0}}\right)^{2}}.
\end{equation}
We can see $\varphi$ decreases with increasing magnetic field,
however the second critical field is infinity like the critical
temperature. Superconductor phase exists at any magnetic field.
The absence of a phase transition to the normal state with
increasing magnetic field can be explained as follows. In
Ginzburg-Landau theory transition to normal state takes place when
average distance between vortexes becomes the order of the
correlation length $\xi$. A line in center of a vortex is normal.
If distance between the centers of vortexes is $\xi$ hence the
system is divided into the superconducting regions size of $\xi$.
However the correlation length is size of a normal phase nucleus
arising fluctuationally in the superconductor. Thus the
fluctuations destroy superconducting phase if distance between
centers of vortexes is less than $\xi$. As mentioned above
switching of the field $\upsilon$ changes the situation. The
external pair potential holds superconducting phase regions the
size of $\xi$ Eq.(\ref{3.10}). Thus the superconducting phase is
stable at any concentration of the vortexes hence at any magnetic
field intensity.

Thus the proposed model of hypothetical superconductivity
demonstrates the principal differences from results of BCS and
Ginzburg-Landau theory due presence of the external pair
potential. In a case of decreasing of Cooper pair's energy by the
external field the energy gap tends to zero asymptotically with
increasing temperature. Thus the ratio between the gap and the
critical temperature is $2\Delta/T_{\texttt{C}}=0$ instead of a
finite value in BCS theory. Moreover the energy gap tends to zero
asymptotically with increasing magnetic field. Thus critical
temperature and the second critical magnetic field are equal to
infinity formally. Unlike BCS model the Ginzburg-Landau parameter
is not constant and it increases with temperature. This means that
at large temperature all superconductors in the external pair
field become type II superconductors. However the first critical
magnetic field and maximal current of a thin wire are finite
values and decrease with temperature. This model does not solve
the problem of room-temperature superconductivity, however it
allows to reformulate the problem. Possible practical realization
of the model is proposed in \cite{grig}, where a source of the
external pair potential has been constructed.

\appendix
\section{Simple method of calculation of the second critical field $H_{\texttt{c2}}$ in Ginzburg-Landau
theory.}\label{field}

Let a superconductor is in magnetic field $\textbf{H}\upuparrows
Oz$. It is convenient to choose a calibration $A_{y}=Hx$. Then
Ginzburg-Landau equation has a form \cite{tinh}:
\begin{equation}\label{A1}
    \xi^{2}\left[-\frac{\texttt{d}^{2}}{\texttt{d}x^{2}}+\frac{2\pi i}{\Phi_{0}}Hx\frac{\texttt{d}}{\texttt{d}y}
    +\left(\frac{2\pi
    H}{\Phi_{0}}\right)^{2}x^{2}\right]\varphi-\varphi+|\varphi|^{2}\varphi=0
\end{equation}
Here unlike the standard method we retained a term $\varphi^{3}$.
When the field strength is of about $H_{\texttt{c2}}$
superconductor are pierced by many vortices, so that the order
parameter is strongly nonhomogeneous $\varphi=\varphi(x,y)$, it
varies over distances of the order of the coherence length $\xi$.
We can consider the order parameter is real $\varphi=\varphi^{+}$
and average it over the system so that
$\langle\varphi(x,y)\rangle=\varphi=\texttt{const}>0$. In
addition, we can suppose $x^{2}=\xi^{2}$. Then Eq.(\ref{A1}) takes
the form:
\begin{equation}\label{A2}
    \xi^{4}\left(\frac{2\pi
    H}{\Phi_{0}}\right)^{2}\varphi-\varphi+\varphi^{3}=0,
\end{equation}
and the order parameter is
\begin{equation}\label{A3}
    \varphi=\sqrt{1-\xi^{4}\left(\frac{2\pi H}{\Phi_{0}}\right)^{2}}.
\end{equation}
We can see $\varphi$ decreases with increasing magnetic field. At
the field
\begin{equation}\label{A3}
    H_{\texttt{c2}}=\frac{\Phi_{0}}{2\pi\xi^{2}}=\sqrt{2}\chi H_{\texttt{c}}
\end{equation}
second order phase transition takes place
$\varphi(H_{\texttt{c2}})=0$. At $H>H_{\texttt{c2}}$
superconducting phase is absent.




\begin{thebibliography}{99}

\bibitem{mahan} Gerald D. Mahan, Many-particle physics (Physics of Solids and Liquids), $3^{\texttt{rd}}$ edition, Plenum Publ. Corp.
2000

\bibitem{ginz} V.L. Ginzburg, D.A. Kirzhnits, High-temperature superconductivity,
Consultants Bureau, New York  1982

\bibitem{matt1} R. D. Mattuk, B.Johansson, Advances in Physics \textbf{17},
509 (1968).

\bibitem{sad} M.V. Sadovskii, \emph{Diagrammatics: Lectures on Selected Problems in Condensed Matter Theory}, World Scientific, Singapore 2006.

\bibitem{schr} John R. Schrieffer, Theory of Superconductivity,
benjamin, 1964.

\bibitem{matt2} Richard D. Mattuk, \emph{A guide to Feynman diagrams in the many-body problem}
(H. C. Oersted Institute University of Copenhagen, Denmark, 1967).

\bibitem{tinh} Michael Tinkham, \emph{Introduction to superconductivity}
(McGRAW-HILL Book Company, 1975).

\bibitem{grig} K.V. Grigorishin,  arXiv:1408.6761 [cond-mat.supr-con]
(2014)


\end{thebibliography}
\end{document}